# Reaffirmation of Cosmological Oscillations in the Scale Factor from the Pantheon Compilation of 1048 Type Ia Supernovae

H. I. Ringermacher[1] and L. R. Mead[2]
Dept. of Physics and Astronomy, U. of Southern Mississippi, Hattiesburg, MS 39406, USA
[1]ringerha@gmail.com ; [2]lawrence.mead@usm.edu

We observe damped temporal oscillations in the scale factor at a dominant frequency of ~ 7 cycles/Hubble-time in the Pantheon Compilation of 1048 type Ia supernovae (SNe). The residual oscillations observed in the Pantheon data closely matches and reaffirms our initial observation of oscillations from earlier SNe data (primarily SNLS3, 2011) at 2-sigma confidence. The nearly identical shapes in amplitude, frequency, phase and damping constant makes it highly likely the signal is real. Furthermore, 2/3 of the Pantheon SNe cover different portions of the sky compared with SNLS3 strengthening this conclusion. Our model describing the oscillation, presented in an earlier paper, is a simple scalar field harmonic oscillator coupled to the $\Lambda$CDM Friedmann eqn, but carried into the present epoch. The scalar field energy density plays the role of the dark matter energy density in $\Lambda$CDM cosmology, fits well as an average, and closely matches the present dark matter density parameter, suggesting the oscillation play a role in the dark matter sector. Temporal oscillations in the scale factor and its derivative, as described in the present work, would also induce temporal oscillations of the Hubble parameter.

## 1. INTRODUCTION

Now that cosmology has been reborn as "precision cosmology", astronomers are willing to utilize the data and models to search for deviations from the Concordance Model. Time-dependent cosmological parameters are one possibility. In particular, parameters oscillatory in time have recently been considered and researched (Brownsberger, Stubbs & Scolnic, 2019). These include temporal oscillations in the universal gravitational constant, *G* (Barrow & Parsons, 1997), as well as in dark energy. We comment on this study in the Conclusions. Periodicity with respect to redshift is not considered in this work, nor are periodic redshifts – all distinct from temporal periodicity.

In earlier work (Ringermacher & Mead 2014) we demonstrated a novel method of transforming a standard Hubble diagram of distance modulus vs. redshift into a scale factor vs. cosmological time plot for a data set of type Ia SNe together with radio galaxy beacons. We refer to this data as "CDR" (Conley, et al. 2011, Daly & Djorgovski 2004, Riess et al. 2004). The transformation depends only on the knowledge of spatial curvature and no other properties of a universe model. This process is described in Appendix A. We showed that the $\Lambda$CDM model for zero spatial curvature in the scale factor plot was the best fit at 98% confidence to the CDR data set thus confirming the technique. In a second follow-up paper (Ringermacher & Mead 2015), hereafter R15, we analyzed the scale factor plot seeking an inflection point, which would define the "transition-time" when the universe changed from a decelerating expansion to accelerating. Instead, we observed a number of relative minima, apparently arising from damped oscillations in the scale factor at a dominant frequency of ~ 7 cycles/Hubble-time (7 Hubble-Hertz, or 7HHz , a convenient unit coined in R15, where $1 HHz = 0.1023 h_{100} \, Gyr^{-1}$) thus accounting for the large variation of transition redshifts in the literature.

In this paper we are strictly interested in extracting the oscillations from the observed scale factor data. In doing this we assume, as mentioned earlier, the scale factor follows standard ΛCDM cosmology with fixed cosmological parameters. Thus we are examining perturbations in ΛCDM and not comparing various oscillating models of ΛCDM. All cosmological parameters are fixed as fitted to the data sets according to ΛCDM cosmology

## 2. DATA ANALYSIS - NOISE

2.1. *Random Noise*

The oscillations in R15 were analyzed using a variety of methods including Gaussian smoothing, Fourier analysis, auto-correlation and a careful noise study of 5000 trials generating random noise with the same temporal distribution as the data. The noise study showed that the chance of generating a specific sharp (1 HHz FWHM) frequency from smoothing random noise over a 20 HHz bandwidth was 1/20 while the chance of seeing an injected signal at that frequency with amplitude adjusted for the expected SNR (~0.2) was 1/2. Thus the "effective likelihood" of observing a noise-generated signal from data processing was 1/10.

New SNe data, the "Pantheon compilation" was recently made public (Scolnic et al. 2018). Two thirds of the Pantheon SNe cover different portions of the sky compared with CDR. 390 of the SNLS3 points are duplicated in Pantheon. It is important to note, however, that although there are duplicate data points, the data have been processed differently in each set and the resulting statistical uncertainties are significantly different. The temporal noise distributions are similar but the published uncertainties in modulus for Pantheon are nearly double those of CDR, including the duplicate SNe. Although the CDR and Pantheon data are not independent, the residuals, comprising these new noise statistics are independent. The noise study of R15 thus applies to the present work as well. Therefore the likelihood of finding the same noise-generated signal by smoothing in both the CDR set and the Pantheon compilation is, conservatively, 1/100.

2.2. *Systematic Noise*

Possible sources of systematic noise lie in the data processing procedures, including; (a) transformation to scale factor, and (b) data smoothing. We do not consider systematic uncertainties in the SNe themselves although such uncertainties would have to be temporally coherent to high redshift to emulate our results, which seems unlikely. In all cases, we take the data as given (redshift , distance modulus and RMS modulus error) in the source papers, recognizing that, technically, there is a covariance between distance moduli, though with very weak correlation.
.
Data smoothing is separately discussed in this paper and is shown to be a possible systematic noise source, though an unlikely one. Transforming from a Hubble diagram to scale factor is a monotonic, nonlinear procedure. We expect that this would not introduce oscillatory components. However, to confirm this, we performed a simulation transforming distance modulus vs. redshift to scale factor vs. cosmological time for 2000 random redshifts. The data was processed as described in the section 3.1. A Fourier analysis showed only random noise over the bandwidth of interest.

## 3. PANTHEON DATA ANALYSIS

In the present paper we present the Pantheon data in a standard Hubble diagram, then transform the modulus vs. redshift plot to a scale factor vs. cosmological time plot following the detailed procedure in Ringermacher & Mead, 2014. The Pantheon time domain data are then analyzed for oscillations and compared to an identical reanalysis of the previous CDR data.

Figure 1(left) displays the Pantheon compilation as a standard Hubble diagram of distance modulus vs. redshift for a best-fit Hubble constant of 68.7 km/s/Mpc and SN Ia $M_0 = -19.36$. Redshift, z, is directly transformed to scale factor using $a(t) = 1/(1+z)$.

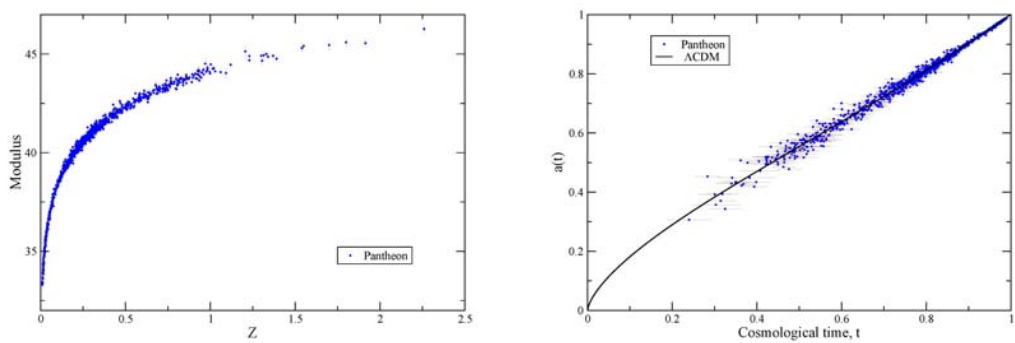

**FIG.1** Distance modulus vs. redshift for Pantheon SN1a data (left). SN abs. mag. $M_0$ = -19.36. Scale factor vs. cosmological time (right). Scale factor and time are normalized to Hubble constant $H_0$ =1. t =1 is z = 0.  t =0.43 corresponds approximately to z = 1. The fitted curve is the $\Lambda$CDM model scale factor.

Modulus is transformed to cosmological time carefully following the definitions and procedure described in (Ringermacher & Mead 2014). Dominant error is in the time direction since redshift error is negligible. The resultant scale factor data plot for a flat 3-space FRW metric is shown on the right. The solid curve is the best fit $\Lambda$CDM model for $\Omega_\Lambda = 0.73$ and $\Omega_m = 0.27$.

### 3.1. Methodology

The data of Figure1 (left) are first binned into 128 bins on the scale of 0 to 1.0 – equal time bins fine enough (0.007813 ) to faithfully resolve waveforms at a Nyquist frequency of 64 HHz. We chose to bin over the full unit time scale for two reasons: (1) working with unit time scale is natural for our model and, (2) equal cosmological time bins are essential for faithful Fourier analysis. Random time errors are effectively transformed to scale factor errors after plotting the ascending time values against scale factor, as described in Appendix A. Once the time values are binned, they are fixed and the effective error moves to the average scale factor for that time bin. To calculate the actual scale factor error from the time error, we would use the Friedmann equations for ΛCDM relating a(t) to t. For $\Omega_\Lambda = 0.73$ and $\Omega_m = 0.27$ (ignoring radiation) there is an exact solution. We seek residual oscillations on the $\dot{a}(t)$ curve to compare with the results of Ringermacher & Mead (2015). This can be derived in three ways: (1) One can fit a polynomial to the binned data and then subtract the fit function (quadratic) leaving the residuals or, (2) one can subtract the ΛCDM model leaving residuals. The residuals are then processed using a wide-baseline derivative to reduce noise (described in Appendix A of Ringermacher & Mead (2015)) followed by Gaussian smoothing to further reduce noise, or (3) Apply a wide-baseline derivative directly to the binned $a(t)$ curve followed by the same Gaussian smoothing. All methods were applied and produced the same results. We describe method (3). An 8 time-bin (0.06 in time or ~ ½ wave period) derivative was used. This is a "two-point" derivative separated by 8 time bins that results in exceptional noise reduction. We demonstrated in R15 that wide-baseline differentiation effectively eliminates differentiation noise normally created by standard 2-point (one time bin) derivatives. The RMS noise reduction of n-bin over 1-bin differentiation goes approximately as n. This was confirmed through a simulation performing wide-baseline differentiation of our damped oscillation including noise. Wide-baseline differentiation of a function $f(t)$ is defined from

$$\dot{f}_i = \frac{f_{i+n/2} - f_{i-n/2}}{t_{i+n/2} - t_{i-n/2}} = \frac{f_{i+n/2} - f_{i-n/2}}{n\Delta t}, \quad (1)$$

where *n* is the number of time bins of fixed width $\Delta t$. The RMS noise following *n*-bin differentiation is given by:

$$\delta \dot{f}_i = \frac{\sqrt{(\delta f_{i-n/2})^2 + (\delta f_{i+n/2})^2}}{n\Delta t} \quad (2)$$

In the present case, $f(t)$ is the binned residuals of $a(t)$. The noise in a(t), $\delta a$, is propagated from the uncertainties in the distance modulus, $\delta\mu$, and is given by

$$\delta a_i = \frac{\ln(10)}{10}\left(1 - a(t)^2\right)\delta\mu. \quad (3)$$

This is derived in Appendix B as $\delta t_i$, and we have used, to good first order approximation, the fact that $a(t) \cong t$ over our range of data from time 0.4 to time 1.0. That is, $\delta t_i \cong \delta a_i$. The $\delta t$ of Eq. (3) are the time error bars in the right side, Fig.1. $\delta f$ is the binned error RMS averages of the $\delta a$. The scale factor, $a(t)$, is defined from the red shift, z, and is given by $a(t) = 1/(1+z)$, where the present time, $t_0$, is at $z = 0$.

A Gaussian smoothing function with a moving 5% time window (0.05 smoothing window), on a scale of 0-1.0, ("ksmooth" from Mathcad™ ) was applied following differentiation. This reduced noise further a factor of 8 and narrowed the frequency spectrum to 10 HHz (-6 db point). Frequencies less than 4 HHz were minimized by heavy low-pass filtering with a Gaussian smoothing time window of 13% and then subtracting the resulting low-pass waveform from the 5%-smoothed waveform. A 13% time window is about one wave period (1/7). This will minimally disturb the 7 HHz signal while significantly reducing lower frequencies. Any heavier filtering reduces the 7 HHz signal. The identical procedure was then applied to the earlier CDR data of Ringermacher & Mead (2015). The reanalysis of the earlier set is crucial because that data was not binned prior to smoothing possibly resulting in some spectral distortion where the low-z data was bunched. Any concern that the differentiation procedure could produce such a signal was mitigated by processing the $a(t)$ curve independently using the same binning and smoothing. The same oscillations were found and are described in Section 3.2.

### 3.2. Results

Figure 2 compares the $\dot{a}(t)$ residual oscillations seen in the CDR and Pantheon data following the procedure described above. The raw data is available (Downloadable data, 2020). The time scales for both data sets are normalized for unit Hubble constant. Since the CDR Hubble constant is 66.0 (least-square fit of CDR modulus to ΛCDM for $\Omega_\Lambda = 0.73$ and $\Omega_m = 0.27$ and $M_0 = -19.19$) and Pantheon is 68.75 (least-square fit of Pantheon modulus to ΛCDM for $\Omega_\Lambda = 0.73$ and $\Omega_m = 0.27$ and $M_0 = -19.40$), in order to compare data, we multiplied the Pantheon time scale by 1.041, the ratio of the two Hubble constants, resulting in a correct phase match. 68 equal-time bins common to the two sets, spanning time from 0.45 to 1.0, were analyzed. The amplitudes, shape, phase, frequency and damping are nearly identical. The average RMS SNR is 2.0 for CDR and 2.2 for Pantheon.

We performed a $\chi^2$ analysis comparing the two data sets based on a Z-score so as to express a direct estimate of goodness of data match. The Z-score is given by $Z = (\chi^2 - df)/\sqrt{2df}$, where $df$ is degrees of freedom. We are comparing all 68 points in common between the two sets from $t = 0.461$ to $t = 0.984$ with no adjustable parameters ($df = 68$). The CDR and Pantheon signal amplitudes were compared on a time scale common to both. We consider the CDR set as the data to compare with the Pantheon "effective model data". We use the average of the CDR and Pantheon errors in our chi-square analysis. We find $\chi^2 = 67.40$ giving $Z = 0.026$. This corresponds to a 2-tailed probability of 96% goodness of match. This is equivalent to a match confidence level of $2.1\sigma$, consistent with the observed average SNR of 2.1. We also calculated the Pearson r-correlation coefficient, given by:

$$r = \frac{\sum X_P X_{CDR}}{\sqrt{\sum X_P^2}\sqrt{\sum X_{CDR}^2}}, \qquad (4)$$

where the numerator is the covariance of the CDR and Pantheon data sets and the denominator is the standard deviations. This is possibly the simplest statistic to compare the 2 sets and gives a correlation coefficient $r = 0.90$.

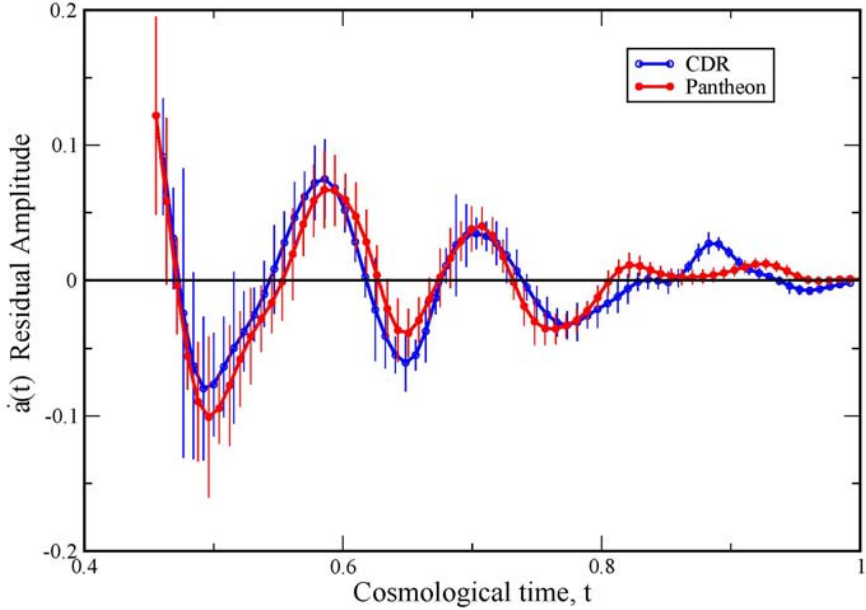

**FIG.2.** Residual oscillations observed in $\dot{a}(t)$ for CDR and Pantheon data are compared.

The Fourier analysis of the waveforms is shown in Figure 3. The dominant frequency is 7.34 ±1.5 HHz in CDR and 7.66 ±1.5 HHz in Pantheon. Lower and higher frequencies are evident, but have been suppressed. The probability of measuring the same frequency (~7 HHz), generated by noise alone, in the two data sets is estimated at 1/100 based on the afore-mentioned noise study. The damping was modeled as an exponential decay, $e^{-\alpha t}$, from time 0.5 to 1.0. The error in the damping constant is dominated by noise near time 0.5 and is the average of upper and lower decay curves. The average least-squares decay constant is $\alpha = 3.7 \pm 0.3$ which closely matches our theoretical model decay constant for $\dot{a}(t)$ oscillations, $\alpha = 3.5 \pm 0.2$, thus strengthening the reality of the observations.

For convenience we separately show correctly scaled $a(t)$ residuals for both Pantheon and CDR in Fig.4 to provide an estimate of the scale factor oscillation amplitude. The a(t) residual signal in Figure 4 also shows the correct phase with respect to the wide-baseline derivative of the residuals calculated in Figure 2, thus validating the wide-baseline derivative and excluding it a possible systematic source of the signal. We also calculated $\chi^2$ and the Pearson-r correlation for the two sets. As before, we used the Z-score approach and found $\chi^2 = 67.40$ ($df = 68$) giving $Z = 0.206$. This corresponds to a 2-tailed probability of 84% goodness of match. We also found a correlation of $r = 0.84$.

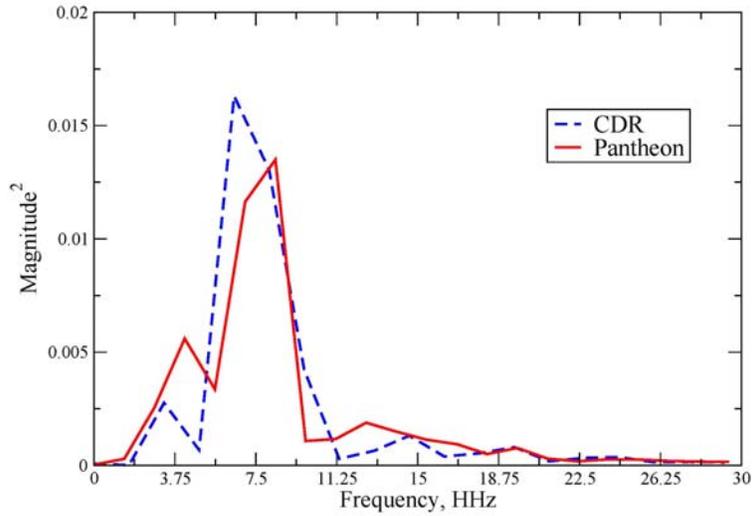

**FIG.3.** Fourier power spectrum of Fig. 2 showing peaks for both data sets at ~ 7HHz. It is also important to note that there may be other frequencies since we have band-passed to extract the dominant frequency.

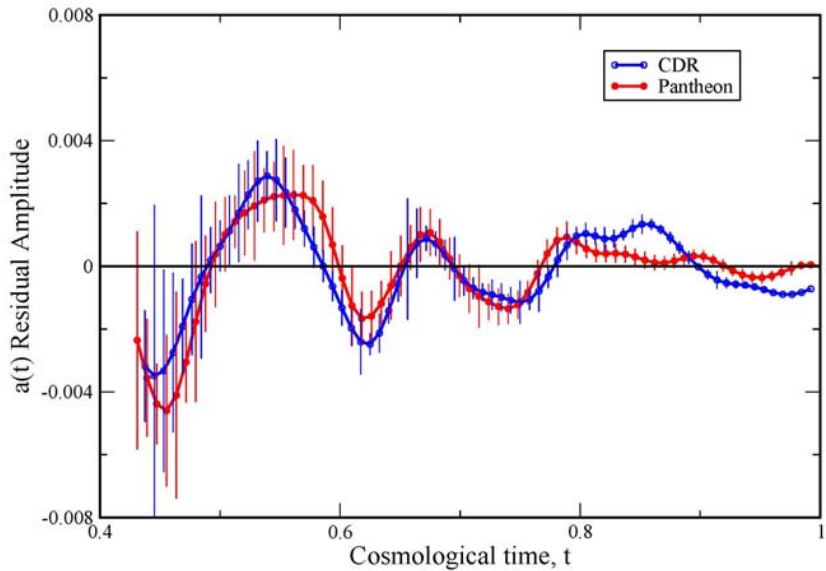

**FIG.4.** Correctly scaled $a(t)$ residuals comparing CDR and Pantheon results.

### 3.2.1 Impact of data redundancy on signals present in 2 data sets

The pantheon data set includes SNe from Conley et al. (SNLS3, part of our CDR data set). We wish to investigate the correlation of the two data sets in the presence of a signal in common. Let us suppose that there is no signal in the recent Pantheon data and that the signal we observe in the earlier CDR is some sort of systematic effect. About 1/3 of the Pantheon SNe were taken from CDR. Those SNe were found to be uniformly distributed from z = 0 (time 0, the present) to z = 1 (time 0.4). We ask what is the effect, on our analysis, of a portion of the CDR signal appearing in Pantheon and how strongly are the two sets then correlated?

To quantitatively answer this question, we performed our smoothing analysis on two simulated binned a(t) residual data sets, each 128 bins as in our analysis. In one set, say representing CDR, we inserted a residual signal (sinusoid resembling our observation) continuously occupying 77 bins from time 0.4 to time 1.0 – as in our analysis. A fraction of those bins, 0.38, or 29 bins uniformly distributed among the 77, containing signal, were incorporated into the second set ( representing Pantheon) at the correct times for those bins. Thus the "Pantheon" set had 29/77 bins with signal plus noise and all remaining bins with zero signal (noise only). We then preformed the same 0.05 time-window smoothing on each data set, the same as in our analysis. The result was the RMS "Pantheon" signal was reduced to 0.38 of the "CDR" signal. That is as expected. For example suppose one of every three points in the Pantheon file of 77 points (roughly our simulation) has unity signal and the remaining two have zero signal. Smoothing will average the three points to give a reduced amplitude over the waveform of 0.33. The Pearson r-correlation for the two simulated files is $r = 0.60$, weak compared to our actual evaluation of the real data, $r = 0.90$. In our real data analysis we do not see such a dramatically reduced amplitude which would have resulted in a SNR of 0.8 instead of our 2.1. Both the correlation and SNR of our actual data are consistent with signal appearing continuously with virtually the same amplitudes in both CDR and Pantheon.

### 4. CONCLUSIONS

We have reaffirmed the presence of temporal oscillations in the scale factor from the Pantheon Compilation of SNe Ia matching our earlier observations (Ringermacher & Mead 2015). This is noteworthy since 2/3 of Pantheon SNe cover new sky compared to the earlier set. The oscillations were observed by converting the standard Hubble diagram of modulus vs. redshift to scale factor vs. cosmological time. The dominant frequency is approximately 7 cycles/Hubble-time, or 7 HHz (Hubble-Hertz), in both sets. In real-time units this is 0.49 Gyr$^{-1}$ for $h_{100} = 0.7$. We reanalyzed our previous CDR data for this comparison in an identical fashion. The average SNR of Pantheon and CDR observed signals is 2.1, consistent with a $\chi^2$ confidence of match at $2.1\sigma$.

Systematic error producing nearly identical temporal effects in 2 data sets seems unlikely. Although the data sets are not independent, the noise residuals are independent (as described in Section 2). We have shown from previous noise studies that the probability of getting the same sharp dominant frequency signal from 2 independent noise distributions is ~1/100. Therefore this statistic applies to the present analysis. We have also performed a simulation showing how the presence of shared data in Pantheon would affect the analysis outcome. This simulation strongly supports the presence of virtually identical continuous signals first observed in CDR and now in Pantheon.

In R15 we proposed a simple harmonic oscillator model with equations of motion similar to those used for inflation and axion theory (Weinberg 2008) but coupled to the Friedmann equations penetrating into the present epoch. This model is summarized in Appendix C. The damping constant of the observed oscillations matches that of our model thus strengthening our conclusions. Though the model follows the observations fairly well in frequency, phase and damping, it is short a factor of two to three in amplitude and so is incomplete. The scalar field energy density in our model substitutes for the dark matter energy density in ΛCDM cosmology, fits well as an oscillating average, and matches the present dark matter density parameter suggesting this oscillation may be related to the dark matter sector.

Brownsberger, et. al (Brownsberger, 2019) searched Pantheon for oscillations in distance modulus against conformal time, as mentioned in the Introduction, but found none claiming a lower limit of 0.036 mag over a wide-band search from 0.056 $Gyr^{-1}$ (0.8 HHz) to 56 $Gyr^{-1}$ (800 HHz). We estimate that our scale factor signal of peak-peak amplitude 0.0028 has an equivalent peak-peak modulus signal of amplitude 0.02 mag. This is ~ ½ their lower limit and thus would not be observable with their wide-band search. We use cosmological time in our analysis.

## APPENDIX A - Construction of Scale Factor plot

### A 1. Theory

We wish to transform from a plot of distance modulus vs. redshift to a plot of scale factor vs. cosmological time. We begin by writing the FRW metric for the ΛCDM model:

$$ds^2 = dt^2 - a(t)^2 \left( \frac{dr^2}{1-kr^2} + r^2 d\theta^2 + r^2 \sin^2\theta d\phi^2 \right) \qquad (A1)$$

We choose a flat 3-space from current measurements and set $k = 0$. We note that $r, \theta, \phi$ are "frozen" or comoving coordinates. However, they define a position for each galaxy observation imagined to span from the present to distant past thus representing a family of red shifts and coordinate distances with an implicit time dependence. The differential light travel times between SNe samples are summed to generate the cosmological time, much like adding rungs on a "time ladder". A formal discussion of this approach is presented below.

Lookback time, $t_L(z)$, is traditionally calculated from the following integral:

$$t_L(z) = t_H \int_0^z \frac{dz'}{(1+z')E(z')} , \qquad (A2)$$

where
$$E(z) = \sqrt{\Omega_m(1+z)^3 + \Omega_k(1+z)^2 + \Omega_\Lambda} \qquad (A3)$$

is the Hubble parameter for ΛCDM and the density parameters are $\Omega_m$ for dark plus baryonic matter, $\Omega_k$ the curvature parameter and $\Omega_\Lambda$, the dark energy density parameter. In this paper we set $\Omega_k = 0$ for a flat universe. $t_H$ is the present Hubble time, $1/H_0$. We ignore radiation.

Let us examine this formula in detail. The scale factor is defined by $a(t) = 1/(1+z)$. Also we must have, by definition,

$$E(z) = \frac{\dot{a}(t) t_H}{a(t)} , \qquad (A4)$$

where the overdot is the derivative with respect to light travel time (coordinate time), $t$. Clearly, associated with every observed red shift there must be a light travel time from that source. But from the above definitions alone it is clear that the integral (A2) is simply

$$t_L(z) = \int_{t_0}^{t_z} dt' = t_z - t_0, \qquad (A5)$$

Here, $t_0 = 0$. If we choose $t_0 = 1$ for the present time, then we have the cosmological time,

$$\tau(z) = 1 - t_L(z) = 1 - \int_0^z dt' \qquad (A6)$$

where $t_z$ is the light travel time from the source at red shift z and $\tau$ is the dimensionless cosmological time. From the metric, Eq.(A1), the light travel interval along a fixed line-of-sight is:

$$dt = a(t) dr \qquad (A7)$$

This time interval is interpreted as the light travel time interval between two spatially consecutive SNe sightings of a family of observations. The space between the two observations expands such that the sum over all observations of z is the light travel time, $t_z$, from the most distant source to the nearby one at $a(t_0) = a(1) = 1$. One must also be certain that the intrinsic condition, $\dot{a}(1) = 1$, is also satisfied for a proper plot. It remains to describe the coordinate distance, r, in terms of time. We shall be working with several distance measures. Modulus, $\mu$, is a measure of luminosity distance, $D_L$ (Mpc) and is defined from:

$$\mu = m - M_0 = 5 Log\, D_L + 25 \qquad (A8)$$

The luminosity distance is defined from the comoving distance which is our metric coordinate distance, $r$:

$$D_L = r / a(t) \qquad (A9)$$

Thus,

$$dr = d[a(t) D_L]$$

With distances normalized to the Hubble length, and time to Hubble time, our coordinate distance $r(t)$ is the same as the "dimensionless coordinate distance", $Y(z)$, of Daly and Djorgovski ( 2003, 2004) and we may write, adopting their notation;

$$dY = d[a(t)\frac{D_L}{D_H}], \quad (A10)$$

where $D_H = c\, t_H$. We shall keep Eq. (A10) in differential form because both $a(t)$ and $D_L$ vary with each SN measurement and we will analyze our data this way, consistent with Eq. (A7).

Finally, from Eqs. (A6) - (A10), we can write for the empirical dimensionless cosmological time, $\tau$ ;

$$\tau = 1 - \int_0^z a(t)\, dY \quad (A11)$$

together with $\quad\quad\quad\quad\quad a(t) = 1/(1+z)$,

thus relating our plot to direct measurements of red shift and luminosity distance, a numerical procedure which will become clear in the next $a(t)$ plot section.

## A 2  Transforming from Hubble diagram to scale factor plot

We first calculate the cosmological time. We will follow Eq. (A7-A10) very closely and will present a table showing a sample calculation. We assume we have the redshift, $z$, and the luminosity distance in Mpc, $D_L$. $D_L$ is calculated from Eq. (A8), given typical modulus data, $\mu$. Table 2 shows a series of measurements sorted by ascending $z$ in column 1. Shown are a SNe sample set, from Pantheon, starting with the lowest z values followed by a gap jumping to around z = 1 in order to show the changes in the running sum over column 6 to get the cosmological time in column 7. The labeled columns are calculated as follows:

Column 0:   SN name
Column 1:   $z$ , given
Column 2:   $\mu$ , given
Column 3:   $a = 1/(1+z)$
Column 4:   $D_L$ in Mpc from Eq.(A8) given modulus $\mu$
Column 5:   $Y = a\dfrac{D_L}{D_H}$, $\;D_H = c/H_0 = 4282.743\,\text{Mpc}\;$ ($H_0 = 70.0\,\text{km/s/Mpc}$)
Column 6:   $a \cdot deltaY_i = a_i \cdot (Y_i - Y_{i-1})$
Column 7:   $\tau_j = 1 - \sum_{i=1}^{j} a_i \cdot (Y_i - Y_{i-1})$
Column 8:   $\tau corr_j = \tau_j - 0.009579$

**Table 1: Sample calculation of cosmological time**

| #name | z | $\mu$ | a | $D_L$, Mpc | Y | a*delta Y | $\tau$ | $\tau$ corr |
|---|---|---|---|---|---|---|---|---|
| 2002cr | 0.01012 | 33.26745 | 0.989981 | 45.02876 | 0.010409 | | | |
| 2002dp | 0.01038 | 33.4096 | 0.989727 | 48.07508 | 0.01111 | 0.000694 | 0.999306 | 0.989727 |
| 2009an | 0.01043 | 33.2848 | 0.989678 | 45.38998 | 0.010489 | -0.00061 | 0.99992 | 0.990341 |
| 2006bh | 0.01082 | 33.3686 | 0.989296 | 47.17588 | 0.010897 | 0.000404 | 0.999516 | 0.989937 |
| 1998dk | 0.01209 | 33.3391 | 0.988054 | 46.53932 | 0.010737 | -0.00016 | 0.999675 | 0.990096 |
| 2009kk | 0.0122 | 33.8166 | 0.987947 | 57.98558 | 0.013376 | 0.002608 | 0.997067 | 0.987488 |
| 2010Y | 0.01226 | 33.8274 | 0.987888 | 58.27469 | 0.013442 | 6.51E-05 | 0.997002 | 0.987423 |
| * | * | * | * | * | * | * | * | * |
| 06D4cl | 0.99781 | 43.9722 | 0.500548 | 6229.311 | 0.728054 | -0.01525 | 0.476724 | 0.467145 |
| 04D3dd | 1.00279 | 44.4398 | 0.499303 | 7726.094 | 0.900746 | 0.086441 | 0.390284 | 0.380704 |
| SCP05D0 | 1.014 | 44.21445 | 0.496524 | 6964.501 | 0.807437 | -0.04659 | 0.436873 | 0.427294 |
| Eagle | 1.021057 | 44.24235 | 0.494791 | 7054.561 | 0.815022 | 0.003766 | 0.433107 | 0.423528 |
| 04D4dw | 1.029152 | 44.0889 | 0.492817 | 6573.248 | 0.756386 | -0.02901 | 0.46212 | 0.452541 |
| 06D3en | 1.037247 | 44.0473 | 0.490859 | 6448.519 | 0.739085 | -0.00853 | 0.470646 | 0.461067 |
| SCP06C0 | 1.045342 | 43.99835 | 0.488916 | 6304.781 | 0.719751 | -0.00949 | 0.480136 | 0.470557 |

There are several points to note in order to properly calculate cosmological time. Column 6 clearly shows the presence of noise. This is effectively smoothed by the integration in column 7 but cosmological time nevertheless carries the noise. More importantly, two criteria must be satisfied: A; $a(1) = 1$ and B; $\dot{a}(1) = 1$. An inspection of the table at row 2 shows that $a \neq \tau$. That is because this a is not the one for the present time, but rather for the nearest measured z. So there is an apparent time gap of $0.009579$. This is subtracted from $\tau$ to generate $\tau$ corrected in column 8. In effect this gap is an amount $\Delta a(t) \simeq z$ by virtue of the definition of $a(t)$ and is considered an integration constant. Condition A is then satisfied and the data is centered on the present time. $\tau corr$ is used in the final plot but will be referred to as cosmological time. $\tau corr$ is considered ascending cosmological time and is plotted as the abscissa against the ordinate, $a(t)$ in column 3. Thus the random noise present in the cosmological time is transformed to scatter in $a(t)$.

## APPENDIX B – Error Propagation

We derive formula (3), the time error propagated from the error in distance modulus.

Distance modulus as a function of Luminosity distance, $D_L$ (Mpc), is defined from:

$$\mu = \frac{5 \ln(D_L)}{\ln(10)} + 25 \tag{B1}$$

Luminosity distance is defined from the comoving (coordinate) distance, $r$, and scale factor, $a(t)$:

$$D_L = \frac{r}{a(t)}, \quad \text{or,} \quad r = a(t) D_L \tag{B2}$$

From the metric, (A1), the light travel time along a fixed line-of-sight is:

$$dt = a(t) dr \tag{B3}$$

From (B2), $\quad dr = da(t) D_L + a(t) dD_L \tag{B4}$

Thus, from (B4) and (B3), changing differentials to errors, the error in light travel time is:

$$\delta t = a(t) \delta a(t) \frac{D_L}{D_H} + a(t)^2 \frac{\delta D_L}{D_H} \tag{B5}$$

Here we have explicitly normalized to Hubble distances, $D_H = c/H_0$, as in our plot procedure. But $\delta a(t) = a(t)^2 \delta z$ and $\delta z$ is negligible. So, (B5) is well-approximated by:

$$\delta t = a(t)^2 \frac{\delta D_L}{D_H} \tag{B6}$$

From (B1) we have:

$$\delta \mu = \frac{5}{\ln(10)} \left( \frac{\delta D_L}{D_L} \right) \tag{B7}$$

Substituting (B7), the normalized time error, (B6), becomes:

$$\delta t = \frac{\ln(10)}{5} a(t)^2 \left( \frac{D_L}{D_H} \right) \delta \mu \tag{B8}$$

Luminosity distance can also be written as an energy integral (Weinberg, 2008):

$$\frac{D_L}{D_H} = (1+z) \int_0^z dz / \sqrt{\Omega_m (1+z)^3 + \Omega_V} \tag{B9}$$

This is an elliptic integral that we approximate as a polynomial in $z$ for the Omegas given in this paper ($\Omega_V = 0.73$, $\Omega_m = 0.27$):

$$\frac{D_L}{D_H} \cong z + \frac{z^2}{2} \tag{B10}$$

This approximation is valid to within a few % for $z < 2$.

Substituting (B10) into (B8) yields our Eq.(3):

$$\delta t = \frac{\ln(10)}{10} \left(1 - a(t)^2\right) \delta \mu \tag{B11}$$

where $a(t)$ is defined from the redshift as $a(t) = 1/(1+z)$.

## APPENDIX C – Scalar Field Model

Complete details, including Lagrangian, of our scalar field model are presented in R15. The model basically emulates an inflaton-type field entering the present epoch - a simple harmonic scalar field coupled to gravity through the $\Lambda$CDM Friedmann equation for the scale factor, $a(t)$. Radiation is ignored. The equations of motion are:

$$\dot{a}/a = \sqrt{\Omega_\Lambda + \Omega_b/a^3 + \Omega_\phi}\,, \tag{C1}$$

$$\ddot{\phi} + 3\left(\frac{\dot{a}}{a}\right)\dot{\phi} + m^2\phi = 0\,, \tag{C2}$$

where, $\Omega_\Lambda = 0.735$, is the dark energy density; $\Omega_b = 0.043$ is the baryonic matter density; and $\Omega_\phi$ is the scalar field energy density given by,

$$\Omega_\phi = \frac{1}{2}\dot{\phi}^2 + \frac{1}{2}m^2\phi^2\,. \tag{C3}$$

$m$ is the angular frequency of the scalar field, $\phi$, on a time scale of 0-1; $m = 2\pi f_\phi = 21.84$, or, $f_\phi \simeq 3.5$. The frequency of oscillations, $f_a$, in the scale factor, $a(t)$, comes through as twice this, or, $f_a \simeq 7$, that is, 7 cycles over 1 Hubble-time, the observed frequency in this paper. In (C2) $\dot{\phi}(0) = 0$ and $\phi(0) \approx 1.0$ is adjusted precisely so as to satisfy the constraint, $a(1) = 1.0$. There are no other adjustable parameters.

We note that in (C1) $\Omega_\phi$ has replaced the usual dark matter density. Thus in this model the oscillating scalar field is the dark matter.

Our model overall scale factor match to $\Lambda$CDM was shown in R15 to be excellent and deviates significantly only at very early times, $t < 0.2$. Also, frequency, phase and damping of the oscillations are correctly explained. However, the model scale factor oscillation amplitude, for which there appears to be no adjustable parameter, is a factor of two to three smaller than what is observed. So the model is incomplete.

### ACKNOWLEDGEMENTS

We wish to thank Judith Keating Ringermacher for assistance with data manipulation using Excel™.